# Construction of crystal structure prototype database: methods and applications

Chuanxun Su[1,2], Jian Lv[1,3], Quan Li[3], Hui Wang[1], Lijun Zhang[3], Yanchao Wang[1,3] and Yanming Ma[1]

[1] State Key Laboratory of Superhard Materials, Jilin University, Changchun 130012, China
[2] Beijing Computational Science Research Center, Beijing 100084, China
[3] College of Materials Science and Engineering, Jilin University, Changchun 130012, China

E-mail: wyc@calypso.cn and mym@calypso.cn

## Abstract

Crystal structure prototype data have become a useful source of information for materials discovery in the fields of crystallography, chemistry, physics, and materials science. This work reports the development of a robust and efficient method for assessing the similarity of structures on the basis of their interatomic distances. Using this method, we proposed a simple and unambiguous definition of crystal structure prototype based on hierarchical clustering theory, and constructed the Crystal Structure Prototype Database (CSPD) by filtering the known crystallographic structures in a database. With similar method, a program Structure Prototype Analysis Package (SPAP) was developed to remove similar structures in CALYPSO prediction results and extract predicted low energy structures for a separate theoretical structure database. A series of statistics describing the distribution of crystal structure prototypes in the CSPD was compiled to provide an important insight for structure prediction and high-throughput calculations. Illustrative examples of the application of the proposed database are given, including the generation of initial structures for structure prediction and determination of the prototype structure in databases. These examples demonstrate the CSPD to be a generally applicable and useful tool for materials discovery.

# 1. Introduction

Data-driven approaches are playing an increasingly important role in materials discovery, as they can rapidly screen vast amounts of materials and structures to select promising candidates for further investigation. These methods rely on prior crystallographic information and the ability to learn from available data. Considerable progress in determining atomic structures has been made through combining experimental and theoretical approaches, whereby vast amounts of structural data are accumulated and archived in databases such as the Cambridge Crystallographic Data Centre (CCDC) [1], the Protein Data Bank (PDB) [2,3], the Inorganic Crystal Structure Database (ICSD) [4,5], the Pauling File [6], the American Mineralogist Crystal Structure Database [7], and the Crystallography Open Database (COD) [8]. These generally accessible databases offer unprecedented opportunities for data-driven techniques that can accelerate materials discovery and design. However, large amounts of duplicated information are stored in the databases. For example, 33 entries in the COD for NaCl compounds share the same or similar crystal structures. The duplicated structural information significantly hampers materials analysis and discovery. This problem can be overcome to some extent by the assignment of crystal structure prototypes to analyze and classify the crystal structures in the database.

In 1990, Lima-de-Faria *et al*. provided two important definitions, allowing crystal structure prototypes to be classified as *isopointal* or *isoconfigurational* structures [9]. Based on these definitions, Allmann *et al.* introduced structure types into the ICSD [10], and all data sets with atomic coordinates in the Pauling File [6] have been classified into structure prototypes. Affine mapping was proposed by Burzlaff *et al.* and Hundt *et al.* [11,12] as an alternative approach for determining structure prototypes for periodic structures. This method has already been used in several studies such as for ternary oxide compounds discovery [13] and the Materials Project [14]. The structure prototypes are widely used in the areas of machine learning and high-throughput calculation, such as for crystal structure prediction based on data mining [15,16], ternary oxide compounds exploration using machine learning [13], the AFLOWLIB.org consortium (http://www.aflowlib.org/) [17], and the Materials Project (https://www.materialsproject.org/) [14]. Structure prototypes may also have other potential applications. For example, they can be used to generate initial structures for evolutionary structure prediction methods (e.g., the CALYPSO method [18,19]) and to provide high-quality structures for empirical potential fitting. As a result, the systematic gathering and classifying of structure prototypes is a fundamental and important work.

However, the earlier proposed definitions of structure prototypes adopted in the databases are complicated [10], and manual inspection is occasionally necessary to assign them [10]. Therefore, a simpler scheme is required to define the structure prototypes that can automatically classify the structures in a database into the appropriate structure prototype without human intervention. This manuscript proposes several simple and unambiguous criteria for defining the crystal structure prototype based on our developed numerical descriptor of crystal structures. In particular, we programmed an automated tool (in Fortran code) to analyze, classify, and represent large amounts of structural data, and thus constructed the Crystal Structure Prototype Database (CSPD).

The remainder of this paper is organized as follows. Section 2 discusses the method and detailed implementation to construct the CSPD. We examine some statistical data about the CSPD in Section 3. Applications of our proposed CSPD to materials discovery are presented in Section 4. Finally, a summary is provided in Section 5.

## 2. Methods and implementations

As the amount of available crystallographic data grows, the management of structural information becomes more challenging. This can possibly lead to the problem of data overload, which may be solved by the construction of the CSPD, an important big-data solution. The construction of the CSPD first requires definitions of the crystal structure prototypes and the corresponding norms. On the basis of hierarchical clustering theory, a set of simple and unambiguous norms for assigning a crystal structure prototype to the given structures is proposed using information on composition, symmetry, and configuration. Specifically, this includes (i) the same composition type and total number of atoms in the conventional cell, (ii) equivalent crystal symmetry, and (iii) the difference between two structures being below a quantitative predefined tolerance ($d_t$).

These norms provide theoretical grounds to guide the construction of the CSPD. Norm (i) conveniently provides structures of a given composition type, and norm (ii) considers symmetry information. Norm (iii) calculates the difference between two structures using numerical structure descriptors, and uses the $d_t$ threshold to judge whether they are similar or dissimilar. To obtain a reliable numerical distance between structures, we developed an efficient and robust structure descriptor using interatomic distance. In part 1, we present the details of our structure characterization method. Then, the specific procedure of constructing the CSPD is discussed in part 2.

*2.1. Structure descriptor*

The first critical step in the construction of the CSPD is the development of an effective structure descriptor to identify and eliminate redundant structure entries. This can solve the data-overloading problem without the loss of important information. Several successful methods [12,19–22] have been previously developed to fingerprint structures automatically and evaluate their similarity. For example, Willighagen *et al.* developed a method using a radial distribution function that combines atomic coordinates with partial atomic charges [20], Wang *et al.* developed a bond characterization matrix based on structure bond information [19], and Zhu *et al.* proposed a crystal fingerprint considering the local environment of all the atoms in the cell [22].

After investigating the literature on structure descriptors, we suggest six criteria for the structure descriptor: (i) it should be independent of translation, rotation, or the choice of equivalent cell for the structure; (ii) it reflects differences among atomic species; (iii) it is a continuous function with respect to the motion of the atoms in the structure, and can yield a quantitative measure of the degree of similarity between two structures; (iv) it can generally describe materials of varying dimensions (e.g., 0-D, 1-D, 2-D, and 3-D); (v) an explicit value should be provided as a threshold for judging structures as similar or dissimilar; and (vi) the algorithm to calculate the descriptor should be robust and efficient.

Our proposed new structure descriptor, the Coordination Characterization Function (CCF), considers these criteria and identifies the fingerprint of a structure on the basis of its interatomic distances. It can reflect the coordination character of a structure within a given cutoff radius. In fact, interatomic distances are good structure identifiers because they naturally satisfy criterion (i) and can be easily calculated at low computational cost for different systems, thereby meeting criteria (iv) and (vi). Interatomic distances are established basic information for the structure descriptors of Radial Distribution Functions (RDFs). [20,21,23] However, it should be emphasized that our proposed structure descriptor is significantly different from RDFs, in that it better considers the long-range character of structures owing to it employing the same weighting for a large range of interatomic distances.

The CCF can be obtained by equation (1), and the Gaussian kernel was used to smooth the discrete interatomic distances into continuous functions. To satisfy criterion (ii), a matrix involving different atomic types was constructed to describe the structure. Specifically, each matrix element is related to components of the CCF coming from different pairs of atomic types $i$–$j$, so that

$$ccf_{ij}(r) = \begin{cases} \dfrac{1}{N}\sum_{n_i}\sum_{n_j} f(r_{n_i n_j})\sqrt{\dfrac{a_{pw}}{\pi}}\exp[-a_{pw}(r-r_{n_i n_j})^2], & (i \neq j), \\ \dfrac{1}{2N}\sum_{n_i}\sum_{n_j} f(r_{n_i n_j})\sqrt{\dfrac{a_{pw}}{\pi}}\exp[-a_{pw}(r-r_{n_i n_j})^2], & (i = j), \end{cases} \quad (1)$$

where $n_i$ runs over all the atoms of the $i$-th type within the cell, and $n_j$ runs over all the atoms of the $j$-th type within the extended cell, $r_{n_i n_j}$ is the interatomic distance less than the cutoff radius $r_{cut}$ (usually 9.0 Å), $f(r_{n_i n_j})$ is the weighting function for different interatomic distances, $N$ is the number of atoms in the cell, $a_{pw}$ (usually 60.0 Å$^{-2}$) is an empirical parameter that controls the peak width of the Gaussian function, and $\sqrt{\dfrac{a_{pw}}{\pi}}\exp[-a_{pw}(r-r_{n_i n_j})^2]$ is the normalized Gaussian function that satisfies the condition

$$\int_{-\infty}^{\infty}\sqrt{\dfrac{a_{pw}}{\pi}}\exp[-a_{pw}(r-r_{n_i n_j})^2]dr = 1. \quad (2)$$

In equation (1), when $i = j$, we introduce a factor of 1/2 to avoid over-counting interatomic distances of the same atomic type. Note that a sufficiently large extended cell should be chosen to include all the atom pairs within the cutoff radius $r_{cut}$. Furthermore, a weighting function, $f(x)$, was introduced to avoid discontinuities in the CCFs when an atom enters or leaves the sphere,

$$f(x) = \begin{cases} 1, & (x < r_{cut} - l), \\ \exp[-\dfrac{\alpha(x - r_{cut} + l)}{l}], & (x \geq r_{cut} - l). \end{cases} \quad (3)$$

Note that the shape of $f(x)$ depends entirely on the parameters $\alpha$ and $l$. Parameter $\alpha$ is a decay factor: a large value allows quick decay (figure 1(a)). Parameter $l$ controls the point where $f(x)$ starts to decay (Figure 1(b)). In our studies, $\alpha$ and $l$ are set to 3.0 and 0.5 Å, respectively.

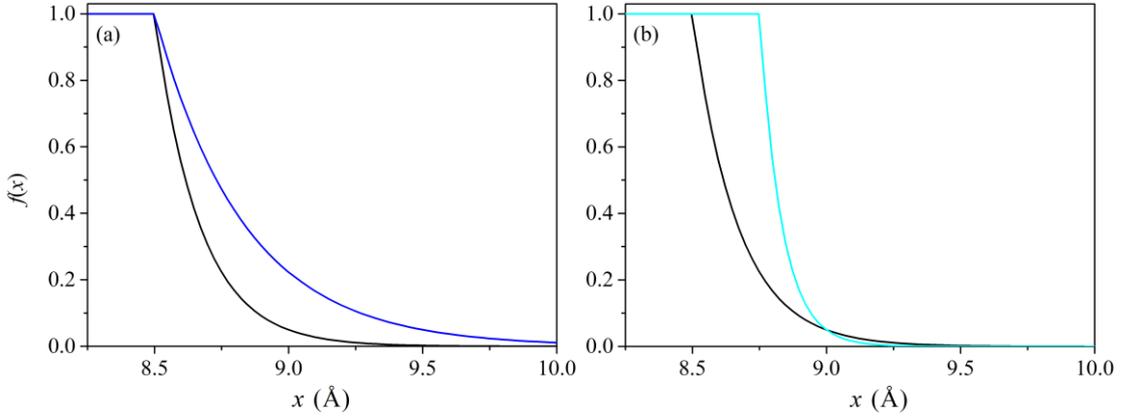

**Figure 1.** The weighting function, $f(x)$, depends on the parameters $\alpha$ and $l$. (a) Weighting functions for $\alpha = 3.0$ (black) and $\alpha = 1.5$ (blue) for fixed $l = 0.50$ Å. (b) Weighting functions for $l = 0.50$ Å (black) and $l = 0.25$ Å (cyan) for fixed $\alpha = 3.0$.

The matrix to describe a structure containing $Nt$ atomic types can be given as

$$M = \begin{bmatrix} ccf_{11} & ccf_{12} & \cdots & ccf_{1Nt} \\ ccf_{21} & ccf_{22} & \cdots & ccf_{2Nt} \\ \vdots & \vdots & & \vdots \\ ccf_{Nt1} & ccf_{Nt2} & \cdots & ccf_{NtNt} \end{bmatrix}. \quad (4)$$

Given that $ccf_{ij}(r) = ccf_{ij}(r)$ can be deduced from equation (1), only the upper triangular matrix is required to characterize the structure. Obviously, a comparison between structures within our scheme then reduces to a comparison of the upper triangular matrix. We employ the Pearson correlation coefficient to measure the degree of similarity between two matrices. It is calculated for discrete points $r_k$ within the range of 0.0–10.0 Å, so that

$$R_{ij} = \frac{\sum_{r_k}[ccf_{ij}^{A}(r_k) - \overline{ccf_{ij}^{A}}][ccf_{ij}^{B}(r_k) - \overline{ccf_{ij}^{B}}]}{\sqrt{\sum_{r_k}[ccf_{ij}^{A}(r_k) - \overline{ccf_{ij}^{A}}]^2 \cdot \sum_{r_k}[ccf_{ij}^{B}(r_k) - \overline{ccf_{ij}^{B}}]^2}}, \quad (5)$$

where $R_{ij}$ denotes the Pearson correlation coefficient; $ccf_{ij}^{A}(r_k)$ and $ccf_{ij}^{B}(r_k)$ represent CCFs from structures A and B, respectively; and $\overline{ccf_{ij}^{A}}$ and $\overline{ccf_{ij}^{B}}$ are the average values of the CCFs for each respective structure. Therefore, the distance for two CCFs can be defined from their correlation coefficient as

$$d_{ij} = 1 - R_{ij}. \quad (6)$$

The summation of $d_{ij}$ with different weighting coefficients yields the distance between the structures, so that

$$d = \frac{\sum_{i=1}^{Nt}\sum_{j=i}^{Nt} c_{ij} d_{ij}}{\sum_{i=1}^{Nt}\sum_{j=i}^{Nt} c_{ij}}. \quad (7)$$

We adopt the following formula to calculate the weighting coefficient $c_{ij}$ for $d_{ij}$ related to atoms of type $i$ and $j$:

$$c_{ij} = \frac{1}{2}\sum_{r_k}[ccf_{ij}^{A}(r_k) + ccf_{ij}^{B}(r_k)]\Delta, \quad (8)$$

where $\Delta$ (usually 0.025 Å) is the step length for $r_k$. Considering that the Pearson correlation coefficient takes values from −1 to 1, the distance between two structures ranges from 0 to 2.

To evaluate the performance of our structure descriptor, it was applied to illustrative examples including TaPd$_2$ and PtK$_2$Cl$_4$. We calculated the CCFs for the conventional cell and the primitive cell of TaPd$_2$, which crystallizes in a MoPt$_2$-type structure (space group *Immm*). [24] Figure 2(a) shows that both CCFs are the same, and the distance calculated by equation (7) is 0.000. Obviously, our structure descriptor satisfies criterion (i). As a further test we considered PtK$_2$Cl$_4$, which has *P4/mmm* (experimental) [25] and *Pnma* (hypothetical) structures, and calculated the CCFs for both structures. The *Pnma* structure is adjusted to the same atom number density as the *P4/mmm* structure. Figure 2(b) shows that the CCFs of the two structures differ greatly, with a distance of 0.655. The value of distance is an effective measure of the degree of dissimilarity between the two structures. Another test examined whether our structure descriptor satisfies criterion (iii). We optimized a metastable structure of CuInS$_2$ (*Pbam*) with four formula units (f.u.). The optimization starts from a structure that lies on the slope of the potential energy surface. After 25 ionic steps, the structure reached the equilibrium point. We calculated ΔE and structural distances between the optimized structure and all the intermediate structures for each ionic step (figure 3).

The structure closest to the equilibrium point tends to have the smallest ΔE and distance values. This test shows that our structure descriptor can quantitatively describe the path taken during structural optimization.

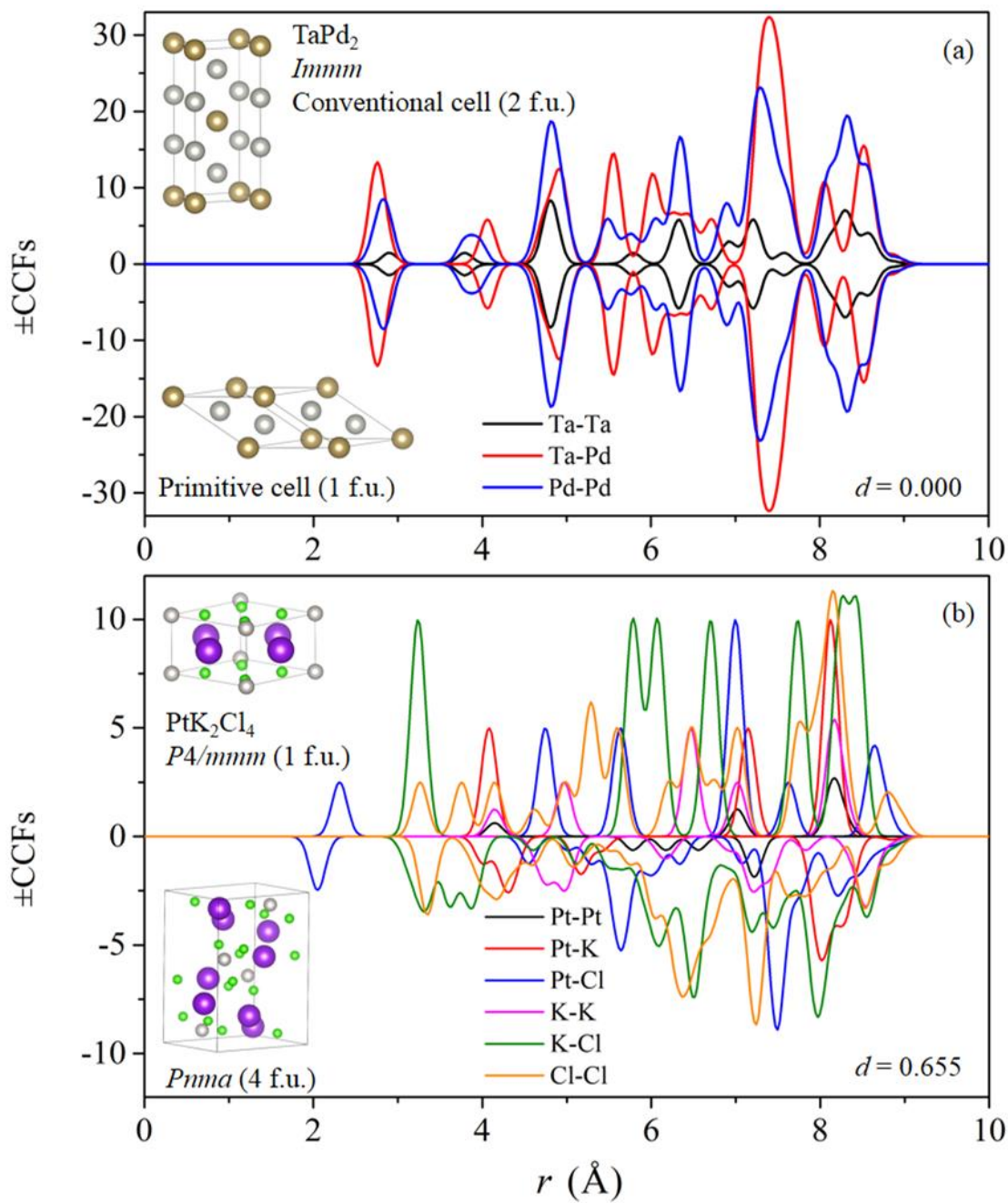

**Figure 2.** Comparisons of the CCFs for (a) TaPd$_2$ and (b) PtK$_2$Cl$_4$. For ease of comparison, the plots show -CCFs for the primitive cell of TaPd$_2$ and the *Pnma* structure of PtK$_2$Cl$_4$.

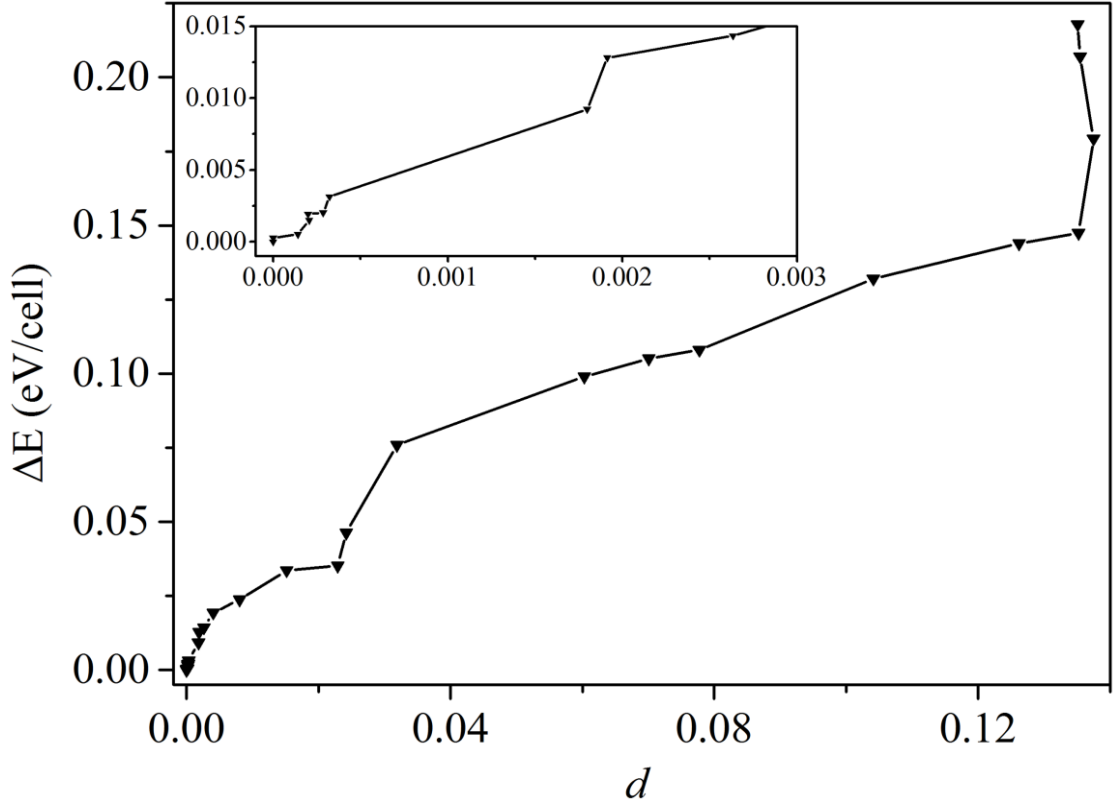

**Figure 3.** Calculated ΔE and structural distances between the optimized structure and all intermediate structures in each ionic step. Inset shows the detailed optimization procedure approaching the equilibrium point.

To determine the threshold for classifying structures as similar or dissimilar, we chose to examine the distances between the structures of $LaC_2$ [26] and $CaF_2$ as an example. Structure prediction runs found 1101 and 385 structures for each, respectively. In each system, two distinct structures were selected as reference structures to calculate the distance to all other structures. Figure 4 shows the calculated distances sorted in ascending order. We have checked the similarity between the reference structures and all the structures of the prediction results via manual inspection. In each test as shown by figure 4, all the structures below the red dashed line are similar with the reference structure, and all the structures above the red dashed line are distinct from the reference structure. Obvious gaps exist between the distances in all four tests. Given these results, we recommend 0.075 as the threshold to determine whether structures are similar. However, this threshold may fail to distinguish large cells with different stacking sequences. We have built two hypothetical structures of element Zn with stacking sequences of ABABCBC and ABABCAB in hexagonal lattice system. The distance calculated by our structure descriptor is 0.00128. The difference of stacking sequence can be reflected by our structure descriptor, but the distance is much smaller than the threshold. Because these two structures are very similar with each other, it is reasonable for the CCF to yield a small distance.

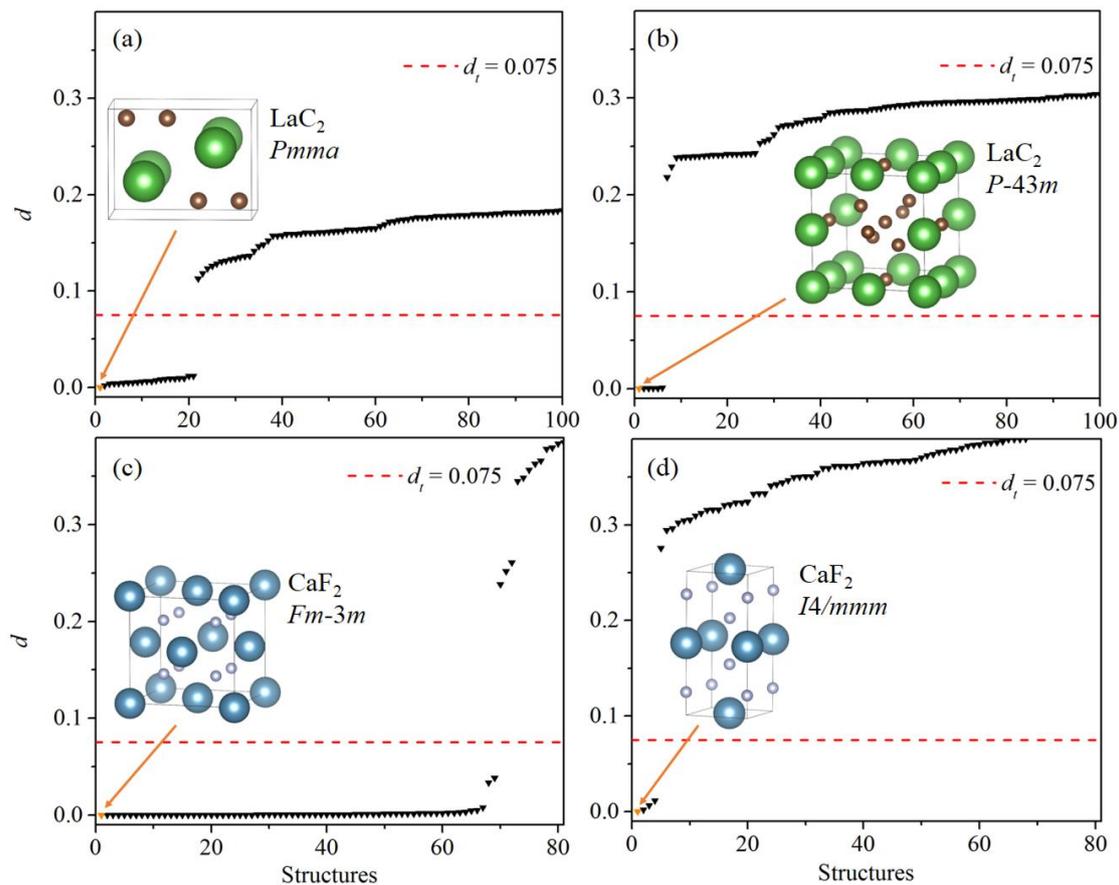

**Figure 4.** Calculated distances of the predicted structures of LaC$_2$ relative to the (a) *Pmma* and (b) *P-43m* structures, and distances for CaF$_2$ relative to the (c) *Fm-3m* and (d) *I4/mmm* structures. All the distances are sorted in ascending order.

## 2.2. Construction of the CSPD

We constructed the CSPD by extracting the crystal structure prototypes from crystal structure database. The norms for assessing the crystal structure prototype were used to classify structures in the COD [8], an open-access collection of crystal structures of organic, inorganic, metal–organic compounds, and minerals, excluding biopolymers. Note that the framework of the CSPD can easily consider crystal structures from multiple databases. The COD has aggregated more than 360,000 entries. All registered users can deposit published or unpublished structures as personal communications or pre-publication depositions at http://www.crystallography.net/cod/index.php. Therefore, all of the structures in the CSPD are freely obtained from the COD.

The procedure of constructing the CSPD comprises four main steps:
(1) filtering structures in the database;
(2) classifying structures according to their composition type and total number of atoms in the conventional cell;
(3) discriminating between inorganic and organic structures;
(4) assessing structural similarity.

Step 1: Filtering structures in the database.

We downloaded the entirety of the crystallographic data of the COD in Crystallographic Interchange File/Framework (CIF) format [27]. Given that the POSCAR format adopted in Vienna Ab-initio Simulation Package (VASP) [28] is convenient for the program to read geometrical information, the CSPD also stores crystallographic data in POSCAR format. Here, the CIF2Cell program [29] was used to transform the crystallographic data of the structures to POSCAR format for the conventional cell. To ensure the CSPD contains only justified structures, all the structures should be filtered: those with incorrect chemical symbols or interatomic distances below 0.6 Å were discarded. Note that the structures in POSCAR and CIF format are stored separately in different directories in the CSPD.

Step 2: Classifying structures according to composition and total number of atoms in the conventional cell.

We selected one specific structure (http://www.crystallography.net/7153176.cif) to demonstrate the determination of composition type. The structure has 46 C, 44 H, 4 N, and 6 O atoms. These numbers, sorted in ascending order, should be denoted as (4, 6, 44, 46); their highest common factor should then be calculated to give the "formula unit". The composition type is then defined as the set of numbers resulting from dividing the original numbers of atoms by the formula unit. The formula unit and composition type can be denoted as, for example, 2 (2, 3, 22, 23). As shown in figure 5, these numbers are used to indicate the file path and file name of the structure in the CSPD. This framework provides convenient access to the structures of a desired composition type in the CSPD.

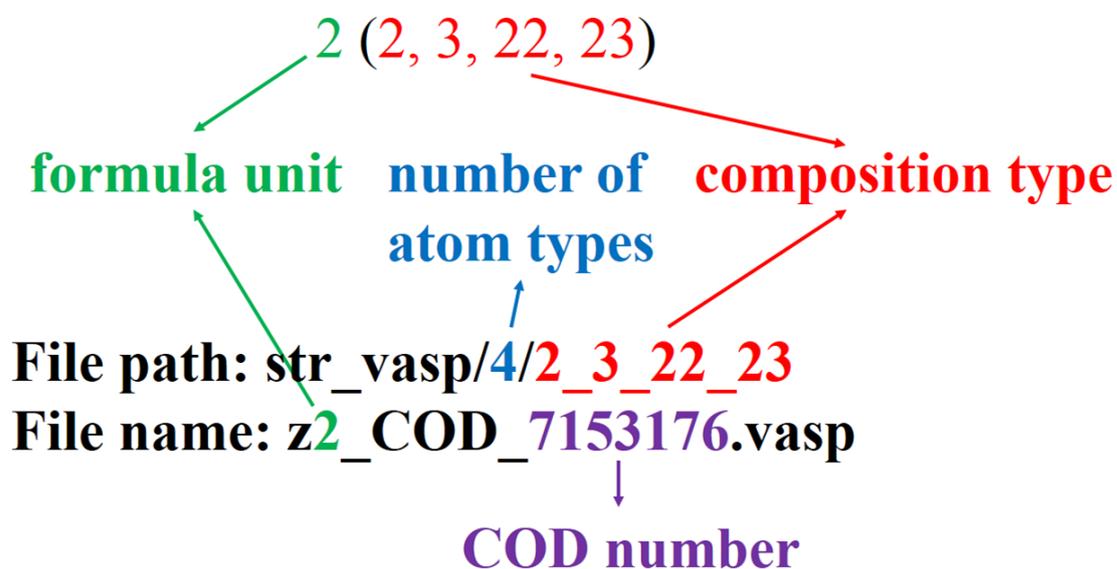

**Figure 5.** Example of the file path and file name of a structure in the CSPD.

Step 3: Discriminating between inorganic and organic structures.

We developed a module in our program to automatically distinguish inorganic and organic structures. Generally speaking, structures containing both C–H and C–C bonds were treated as organic structures. Structures containing C and H had the C–C and C–H interatomic distances calculated. If these distances are less than 1.65 and 1.20 Å, respectively, the structure is recognized as organic. The inorganic and organic structures are stored in different directories in the CSPD.

Step 4: Assessing structural similarity.

To remove multiple entries of similar structures from a database, structural similarity should be assessed and compared with the structures in the CSPD to determine a structure's prototype. We only compared the similarities between structures of the same composition, with the same total number of atoms in the conventional cell. Figure 6 shows a flow chart of the procedure used to compare structural similarity. Specifically, the space groups of the two structures are compared first; if they are the same, the distances between the structures are calculated by our structure descriptor. If the distance is less than 0.075, they are identified as belonging to the same cluster. The CSPD preserves only one structure per cluster.

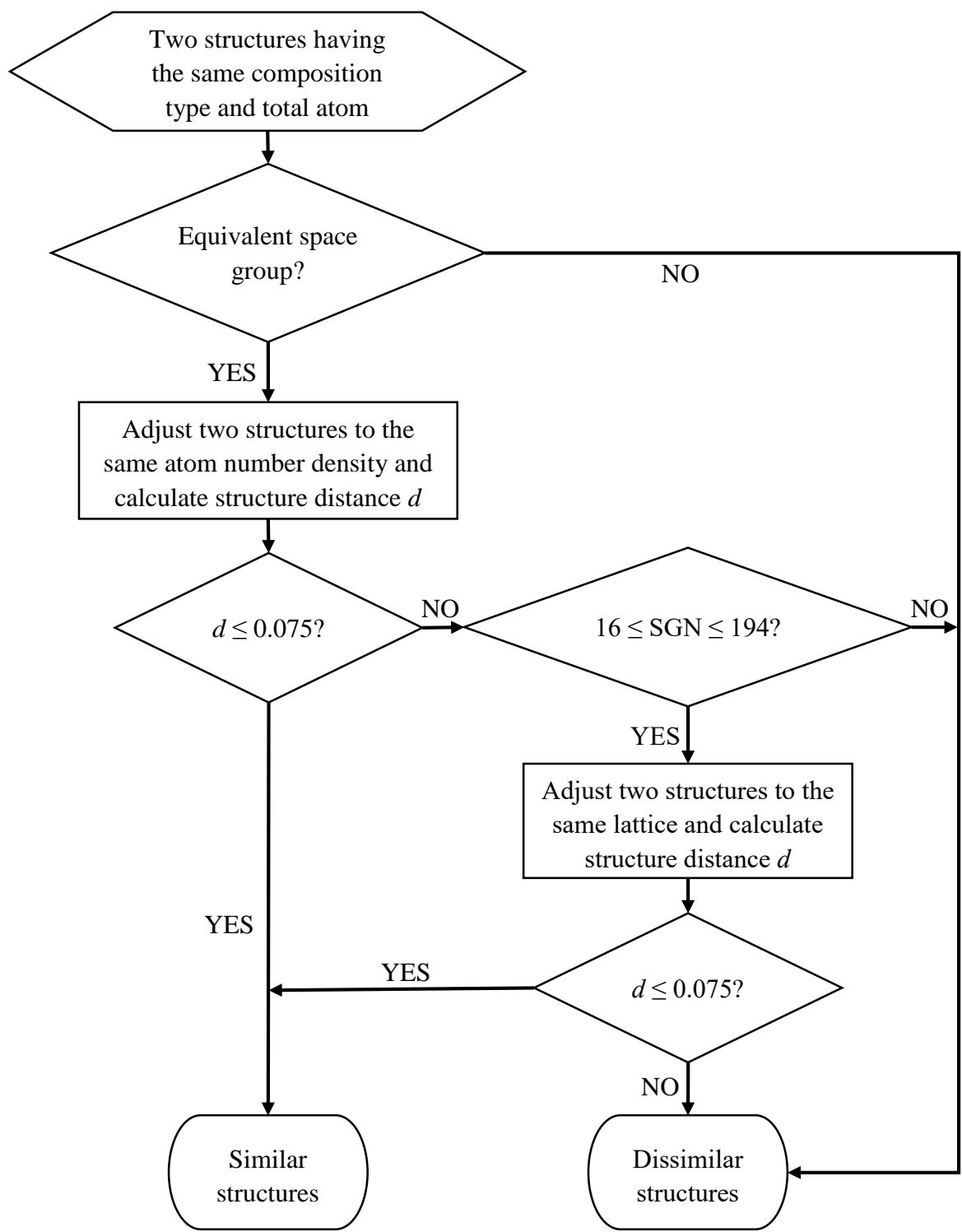

**Figure 6.** Flow chart of the procedure used to compare structural similarity. SGN denotes the space-group number.

Based on this method, we developed a program named the Structure Prototype Analysis Package (SPAP) to analyze the results of CALYPSO structure prediction. CALYPSO is a state-of-the-art structure prediction method and eponymous computer software. It requires only the chemical composition of a given compound to predict its stable or metastable structures at given external conditions (e.g., pressure). It can thus be used to predict or determine the crystal structure. [30–32] CALYPSO calculations typically generate a large number of candidate structures. In particular, many different initial structures lying on the slope of the free-energy surfaces in the same valley reach the same local minimum after local structure optimizations, resulting in many similar structures in the structure searches. The SPAP is a versatile tool for automatic structure analysis; it can be used to evaluate the similarity of the optimized structures and to extract distinct structure prototypes. The SPAP can analyze CALYPSO prediction results for clusters, layered materials, and solids. Furthermore, it can classify the predicted structures and extract the low-energy structures to construct a theoretical structure database. This database can be applied to generate structures for structure prediction. By taking advantage of prior knowledge, this method may improve the efficiency of structure prediction.

## 3. Statistical data of the CSPD

It is convenient to count the number of structure prototypes for each composition type within the framework of the CSPD. The numbers of structure prototypes for each composition type for inorganic and organic systems are listed in Tables I and II, respectively. Table 1 shows that there are 8097 composition types and 26,023 structure prototypes for inorganic systems in our CSPD. Ternary compounds have the most structure prototypes (8098), while quinary compounds have the most composition types (2393) in the inorganic part of the CSPD. Table 2 shows 104,876 composition types and 171,929 structure prototypes for organic systems. The number of structure prototypes for each composition type (NSPC) is given in the fourth column of both tables. It generally decreases with increasing number of atomic types for both inorganic and organic systems. This indicates many as yet unknown structure prototypes for compounds having more atomic types.

**Table 1.** Statistics of inorganic structures in the CSPD. NSPC denotes the number of structure prototypes per composition type.

| Number of atomic types | Number of composition types | Number of structure prototypes | NSPC |
| --- | --- | --- | --- |
| 1 | 1 | 233 | 233 |
| 2 | 252 | 2727 | 10.821 |
| 3 | 1121 | 8098 | 7.224 |
| 4 | 2128 | 7963 | 3.742 |
| 5 | 2393 | 4385 | 1.832 |
| 6 | 1559 | 1940 | 1.244 |
| 7 | 516 | 538 | 1.043 |
| 8 | 108 | 118 | 1.093 |
| 9 | 18 | 20 | 1.111 |
| 10 | 1 | 1 | 1 |

|              | Total | 8097 | 26,023 | — |
| --- | --- | --- | --- | --- |

**Table 2.** Statistics of organic structures in the CSPD.

| Number of atomic types | Number of composition types | Number of structure prototypes | NSPC |
| --- | --- | --- | --- |
| 2 | 198 | 653 | 3.298 |
| 3 | 3222 | 9994 | 3.102 |
| 4 | 13,618 | 33,233 | 2.440 |
| 5 | 30,837 | 57,566 | 1.867 |
| 6 | 32,510 | 44,254 | 1.361 |
| 7 | 17,749 | 19,319 | 1.088 |
| 8 | 5344 | 5494 | 1.028 |
| 9 | 1218 | 1236 | 1.015 |
| 10 | 165 | 165 | 1 |
| 11 | 15 | 15 | 1 |
| Total | 104,876 | 171,929 | — |

Our studies focus on the inorganic systems of the CSPD. The database currently has 233 structure prototypes for elemental solids (see table 1). Table 3 lists the distribution of structure prototypes for different formula units (less than 31). Elemental solids most commonly have two, four, or eight formula units per cell. For binary inorganic systems, the CSPD has 252 composition types and 2727 structure prototypes (see table 1). The top eight popular binary composition types and the corresponding numbers of structure prototypes are listed in table 4. Compounds of the type $AB_2$ have the largest number of structure prototypes among all the composition types in the CSPD. The inorganic structure prototype distribution for the 11 most common ternary composition types is given in table 5. Of these, $ABC_3$, $ABC_2$, and $ABC$ are the three most favorable for inorganic ternary compounds. Inorganic systems of the type $AB_2$ most commonly have two, four, or eight formula units per cell (table 6). This is also commonly seen for other composition types and may be related to the extensive existence of two-fold and four-fold symmetry axes in inorganic systems. [33]

**Table 3.** Distribution of structure prototypes for different formula units (less than 31) of elemental solids. NSP denotes the number of structure prototypes.

| Formula unit | 1 | 2 | 3 | 4 | 6 | 8 | 9 | 10 | 12 | 16 | 18 | 20 | 24 | 28 | 30 |
| --- | --- | --- | --- | --- | --- | --- | --- | --- | --- | --- | --- | --- | --- | --- | --- |
| NSP | 14 | 36 | 9 | 47 | 7 | 30 | 2 | 4 | 9 | 21 | 1 | 3 | 5 | 3 | 4 |

**Table 4.** The eight most common binary composition types and their corresponding numbers of inorganic structure prototypes.

| Composition type | 1:2 | 1:1 | 1:3 | 2:3 | 1:4 | 3:4 | 2:5 | 1:5 |
|---|---|---|---|---|---|---|---|---|
| NSP | 634 | 459 | 266 | 184 | 116 | 100 | 63 | 53 |

**Table 5.** The 11 most common ternary composition types and their corresponding numbers of inorganic structure prototypes.

| Composition type | 1:1:3 | 1:1:2 | 1:1:1 | 1:1:4 | 1:2:4 | 1:2:2 | 1:2:3 | 1:2:6 | 1:2:5 | 2:2:7 | 1:3:3 |
|---|---|---|---|---|---|---|---|---|---|---|---|
| NSP | 549 | 476 | 404 | 378 | 322 | 305 | 298 | 261 | 190 | 125 | 122 |

**Table 6.** Distribution of structure prototypes for different formula units (less than 37) for $AB_2$ inorganic compounds.

| Formula unit | 1 | 2 | 3 | 4 | 5 | 6 | 7 | 8 | 9 | 10 | 12 | 13 | 15 | 16 | 18 | 24 | 27 | 32 | 36 |
|---|---|---|---|---|---|---|---|---|---|---|---|---|---|---|---|---|---|---|---|
| NSP | 19 | 75 | 25 | 213 | 1 | 16 | 1 | 100 | 3 | 2 | 29 | 1 | 2 | 39 | 7 | 20 | 3 | 26 | 10 |

Data abstraction is a key element of data-mining crystal-structure prediction. For example, several established methods for predicting structures are restricted to particular space groups based on the statistics of symmetry of structures. [34] However, our statistical data for the composition types and formula units of structures in the CSPD will assist in materials discovery. We provide two files containing the structure prototype number for each composition type for inorganic and organic systems. Our statistics suggest that structure predictions or high-throughput calculations may focus primarily on certain favorable composition types (i.e., $AB_2$, $AB$, $AB_3$, $A_2B_3$, $ABC_2$, $ABC$, $ABC_3$, $ABC_2D_6$, $ABCD_4$, and $ABC_2D_4$) or numbers of formula units (two, four, and eight). Experimental works can also aim to synthesize the common composition types.

## 4. Applications and results

Here, we illustrate two typical applications of the proposed CSPD: generating initial structures for structure prediction and determining the prototype for a given structure.

*4.1. CSPD for structure prediction*

After transformation of the raw database to a composition–crystal-structure prototype database, the number of candidate structures in the database can be reduced dramatically. Based on our big-data technique (CSPD), we developed an advanced method named the Big Data Method (BDM) to generate structures for structure prediction. It comprises four main steps, as follows:
(1) selecting structure prototypes in the CSPD for a given targeted composition type and formula unit;

(2) substituting elements for the selected structure prototype;
(3) adjusting the lattice parameters for a given volume;
(4) checking the minimal interatomic distances.

To evaluate the performance of our approach, we generated a series of candidate structures for three typical systems (e.g., $CaF_2$, $KN_3$, and $CuInS_2$) by our proposed BDM and also the random sampling method implemented in the CALYPSO package [18,19]. Structure relaxations and energy calculations were performed in the framework of density functional theory within the all-electron projector-augmented wave (PAW) [35,36] method as implemented in the VASP code [28]. The BDM and the random method with symmetry constraints both generate 204, 94, and 144 structures with four formula units for $CaF_2$, $KN_3$, and $CuInS_2$, respectively. The energies relative to the lowest energy of all the distinct structures were sorted in ascending order, and are presented in figure 7. Note that the structure with the lowest energy for all the systems was generated by the BDM, and the structures generated with the BDM tend to be distributed at lower energy. The average computation times for the structure relaxation of these candidate structures are listed in table 7. The BDM appears clearly more efficient than the random method. For $CuInS_2$, the experimental $I$-42$d$ structure was successfully reproduced by the BDM, and it also predicted two novel, energetically degenerate structures of $P$-42$c$ and $P$-4$m$2. The calculated structural parameters for these three phases are listed in table 8. The energies of the new structures relative to the $I$-42$d$ structure are both only 0.001 eV/atom. Therefore, we believe that the proposed metastable structures of $CuInS_2$ are likely to be seen experimentally.

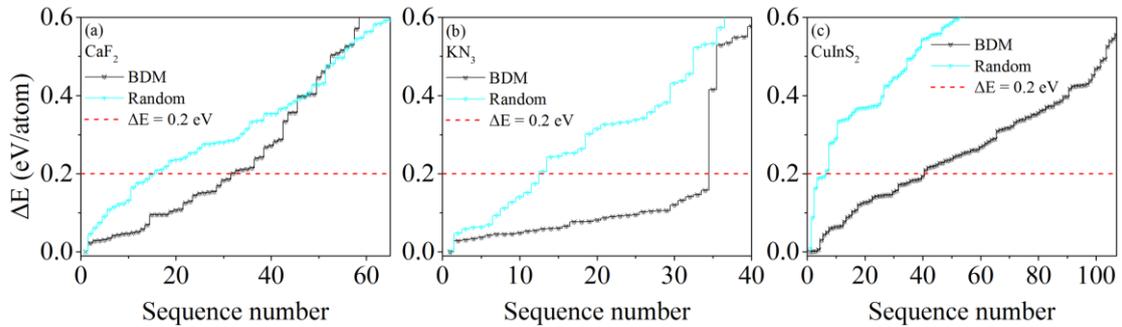

**Figure 7.** Energy stairs of the test results for three systems: (a) $CaF_2$, (b) $KN_3$, and (c) $CuInS_2$.

**Table 7.** Average computational cost for optimizing structures generated by the BDM and the random sampling method.

| System | Method | Computational cost ($s$) |
|---|---|---|
| $CaF_2$ | BDM | 1170 |
| | Random | 2685 |
| $KN_3$ | BDM | 4862 |
| | Random | 7017 |
| $CuInS_2$ | BDM | 1381 |
| | Random | 3327 |

**Table 8.** Optimized structural parameters for the predicted CuInS$_2$ structures.

| Structure | Lattice parameters (Å) | Atoms | Atomic coordination (fractional) | | |
|---|---|---|---|---|---|
| | | | x | y | z |
| *I-42d* | a = 5.5795 | Cu (4a) | 0.5000 | 0.5000 | 0.5000 |
| | c = 11.2362 | In (4b) | 0.5000 | 0.5000 | 0.0000 |
| | | S (8d) | 0.7794 | 0.7500 | 0.1250 |
| *P-42c* | a = 5.5753 | Cu1 (2d) | 0.0000 | 0.5000 | 0.2500 |
| | c = 11.2319 | Cu2 (2e) | 0.0000 | 0.0000 | 0.0000 |
| | | In1 (2f) | 0.5000 | 0.5000 | 0.0000 |
| | | In2 (2b) | 0.5000 | 0.0000 | 0.2500 |
| | | S (8n) | 0.7792 | 0.2504 | 0.1249 |
| *P-4m2* | a = 3.9751 | Cu (1b) | 0.5000 | 0.5000 | 0.0000 |
| | c = 5.5285 | In (1d) | 0.0000 | 0.0000 | 0.5000 |
| | | S (2g) | 0.5000 | 0.0000 | 0.2207 |

The BDM outperforms the random sampling method, because most of the structures in the CSPD were determined by experiment. These structures are more physically justified than those generated by the random method. The initial structures are crucial for efficient structure prediction by global optimization algorithms (e.g., simulated annealing [37,38], minima hoping [39], and evolution algorithm [18,40]). Therefore, the BDM is a powerful approach to generate the initial structures for global structure prediction. Moreover, it can also be used to generate structures for high-throughput calculations.

*4.2. The CSPD for determination of the prototype structure*

The CSPD is valuable for both experimental and theoretical researchers, because it can be used to determine whether a new proposed structure is similar to any known ones. Furthermore, the structure prototype for a given structure can be determined by comparing its structural similarity with the prototype structures in the CSPD. Thanks to our structure descriptor, the structure prototype can be automatically determined by our program SPAP. The procedure comprises two main steps:
(1) screening structures of the same composition type as the given structure in the database;
(2) assessing the distances between the given structure and screened structures in the CSPD.

We took the diamond structure as an example. The distances between the diamond structure and all the structures for other elemental solids with eight f.u. in the databases (i.e., CSPD and COD) were calculated by SPAP. Note that the structures were adjusted to the same atom number density before calculating the distances. The calculated distances were sorted in ascending order, and are plotted in figure 8. According to our norms for crystal structure prototypes, the distance between similar structures must be less than 0.075. In this example, 50 structures in the COD appear under this threshold, all of which are the same diamond structure for C, O, Si, Ge, and

Sn. Only one structure in the CSPD appears under the threshold: the diamond structure. This result further verifies that our crystal structure prototype definition is effective.

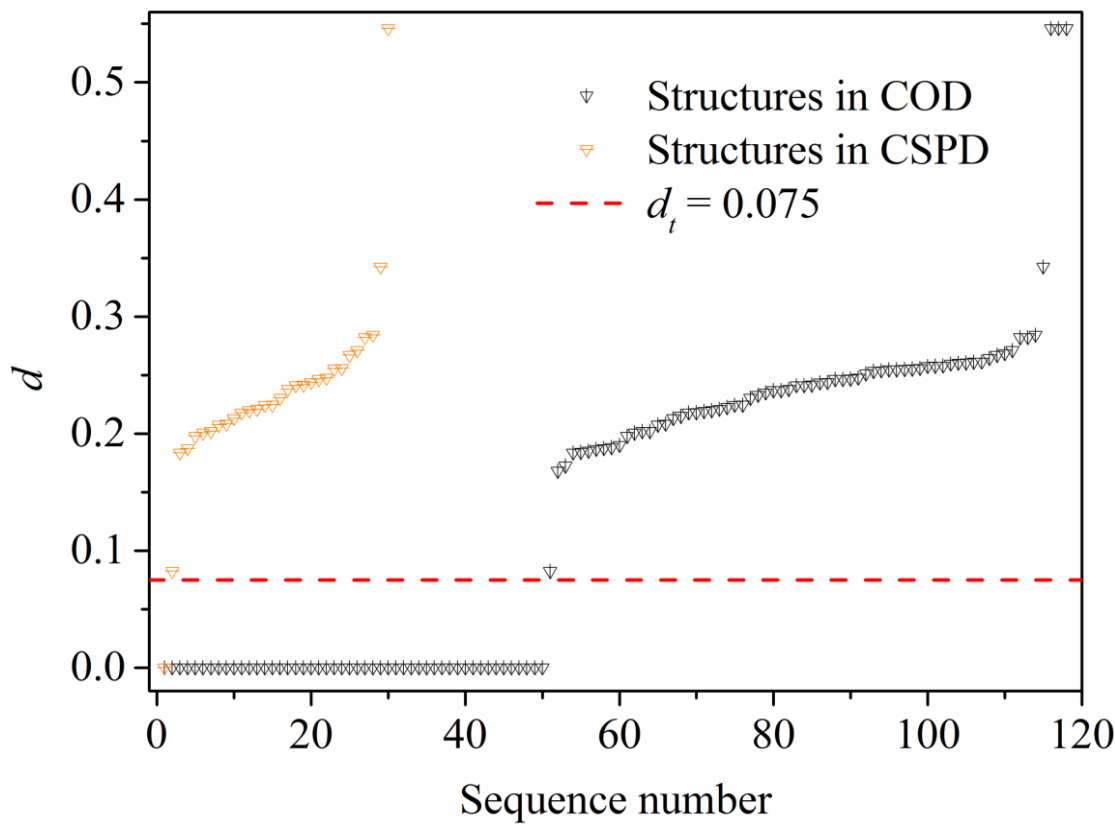

**Figure 8.** Configuration search results for the CSPD and COD.

## 5. Conclusions

In summary, we introduced a set of simple and unambiguous norms for crystal structure prototype determination based on a structure descriptor. They were used as the basis to develop a program for the automatic classification of the structures in a database and to assess the degree of similarity among them. Similar structures were discarded, leaving only a set of unique structures to construct the CSPD. Based on the CSPD, we developed an advanced structure prediction method, the BDM, which was tested using the $CaF_2$, $KN_3$, and $CuInS_2$ systems. The test results show that the BDM outperforms the random sampling method for all three systems. Furthermore, the program SPAP was developed on the basis of the CSPD to determine the structure prototype for a given structure, to remove similar structures from CALYPSO prediction results, and to gather low-energy structures to construct a theoretical structure database, which can be further used to generate structures for structure prediction.

## Acknowledgments

The authors thank the Natural Science Foundation of China for financial support (grants nos. 11274136, 11534003, and 11404128), the National Key Research and Development Program of China under Grant No. 2016YFB0201200, the 2012 Changjiang Scholar program of the Ministry of Education, China Postdoctoral Science Foundation (grant nos. 2015T80294 and 2014M551181), and the fund of CAEP-CSCNS (R2015-03). Some of the computations were carried out at the High Performance Computing Center of Jilin University.